# Exploring Gender-Expansive Categorization Options for Robots

Katie Seaborn
Tokyo Institute of Technology, seaborn.k.aa@m.titech.ac.jp

Peter Pennefather
gDial, inc., p.pennefather@gmail.com

Haruki Kotani
Tokyo Institute of Technology, kotani.h.ab@m.titech.ac.jp

Gender is increasingly being explored as a social characteristic ascribed to robots by people. Yet, research involving social robots that may be gendered tends not to address gender perceptions, such as through pilot studies or manipulation checks. Moreover, research that does address gender perceptions has been limited by a reliance on the human gender binary model of feminine and masculine, prescriptive response options, and/or researcher assumptions and/or ascriptions of participant gendering. In response, we conducted an online pilot categorization study (n=55) wherein we provided gender-expansive response options for rating four robots ranging across four levels of anthropomorphism. Findings indicate that people gender robots in diverse ways, and not necessarily in relation to the gender binary. Additionally, less anthropomorphic robots and the childlike humanoid robot were deemed masculine, while the iconic robot was deemed gender neutral, fluid, and/or ambiguous. We discuss implications for future work on all humanoid robots.

## 1   Introduction

Gender is a social characteristic constructed by people and applied to other agents, especially human-like ones. Indeed, there may be no category as fundamental to human perception as gender [22]. Although the act of gendering is a cognitive process, it has material consequences. Decades of research has shown that people tend to unthinkingly ascribe gender to computer-based agents and react in stereotyped ways to this gendering [21,23–25]. In particular, robots that interact with people have emerged as a target of study. Efforts in human-robot interaction (HRI) have indicated that the perceived gender of the robot [3,4,6,8,10,13,19,27,29], the gender of the participant [8,9,18,30,33], and the gendering of the robot's embodiment in terms of role, task, and/or environment [5,20,31] can play a significant role.

Yet, research on gendering robots remains nascent. The majority of the work conducted so far has been theoretically and methodologically limited in a few key ways. Gender tends to be undertheorized and taken for granted in line with human models of gender [14,34,37], almost exclusively positioned within the gender binary of masculine and feminine. Manipulation checks of gender are underutilized [34]. Researchers may not consider gender, even if gender cues are present in the robot [1]. They may prescribe a limited set of gender options for robots and/or participants or rely on a limited set of gender options when evaluating participant data [1,34,39], typically two: "male/man" and "female/woman." As one critical review using the case study of social robot Pepper found, researchers' models of gender may also affect how participants themselves view the robot's gender as well as how their data is analyzed [34]. What is needed are more rigorous and theoretically grounded explorations of robot gender that are sensitive to inclusive alternatives.

As a first step, we conducted an online pilot study in which participants categorized robots that had varying degrees of anthropomorphism using a *gender-expansive* set of response options. Our research question was: *How does a gender-expansive framing affect participant categorization of robots?* Our contributions are threefold. First, a methodological approach to assessing the gender of social agents using a gender-expansive framework that can be scaled or applied as a manipulation check. Second, empirical findings that show variability across participants when it comes to robot gendering. Third, evidence that anthropomorphism and perceived age of the robot affect how the robot is gendered. We end with a call for the use of gender-expansive response options and greater rigor when approaching gender as a variable of study in social agent research. While the results are preliminary and limited to a passive activity, they illustrate the pervasive and multifaceted nature of the gendering that occurs in—and may be set by—first impressions.

## 2 Theoretical Background

Anthropomorphism is the tendency for people to ascribe human characteristics onto non-human objects, agents, and environments [12]. A wealth of research stretching back decades has pointed to this phenomenon occurring with respect to robots and other agents [21,23–25]. The human tendency to anthropomorphize is thought to be reflexive and largely automatic, where anthropomorphic ascriptions emerge as first impressions [16]. Gender is one such ascription. Gender refers to the social categories by which people organize themselves [17,42]. Gender is typically constructed along a continuum of masculinity and femininity, a model referred to as the *gender binary* [17]. People may move between the poles fluidly, occupy an ambiguous spot in the middle, or reject the binary entirely, i.e., non-binary genders. Sex is often positioned against (and conflated with) gender; by sex, we refer to the biological configurations of the body, including gonads, genitals, hormones, secondary sex characteristics like body hair, and so on [17]. Some have argued that it is difficult to distinguish between gender and sex, at least given the current state of research and societies, arguing for the use of "gender/sex" to account for this [17]. Of course, this may not apply to robots and other mechanical entities, given that machines have no basis in biology. Notably, gender is multi-functional as a precept. It refers to self-identity in terms of inner alignment with gender concepts as well as outward self-expression through bodily accoutrements, i.e., clothing, speech style, stance, etc. It also refers to roles and expectations in society. It can also be ascribed by others, not only to other people but also animals, objects, environments, and non-humans, such as robots.

Gender is not a new concept in research on people's interpretations of robots and other computer-based agents. Moreover, the consensus is that people do gender socially embodied artificial agents, and often in stereotyped ways that align with models of human gender [5,13,21,24,28,29]. For example, Eyssel and Hegel [13] found that people rated the masculine robot as more suitable for masculine tasks, such as transporting goods, over feminine tasks, such as caregiving. However, most research has used limited response options for gendering (or not gendering) robots, in particular based on the human binary model, with perhaps a "no gender" option [34,35]. Yet, as discussed, human gender models are more diverse than the binary. Moreover, robots may not be perceived in line with human gender models, i.e., they may be perceived as having mechanical or robotic genders, or other genders [37]. Finally, research on low and non-anthropomorphic agents, objects, and even shapes and symbols, including birds, fruits, clouds, triangles, and circles, suggests that people will anthropomorphize stimuli in line with gender stereotypes about agency and communality, even in the absence of clear human cues [7,22]. Thus, we may expect that, if given gender expansive options:

**H1:** People will gender robots in diverse ways regardless of the robot's anthropomorphism.

Nevertheless, degree of anthropomorphism may play a role. A face-value judgment of the above literature would lead us to conclude that a robot with highly anthropomorphic cues would stimulate gendering. Consider Ishiguro's Geminoid, a robot modeled after Ishiguro himself through full-body mimicry of his own features, from skin to clothes to hair [26], to the extent that people may not be able to tell Ishiguro apart from his Geminoid. As such, we can hypothesize that:

**H2a:** If the degree of the robot's anthropomorphism is high, people will perceive the robot to have a gender.

At the same time, the absence or scarcity of human gender cues could affect gendering. Søraa [37] conceptualized the notion of "mechanical genders" for robots and other agentic machines. Mechanical genders, which may also be called "robot genders," are particular to robots: a gender construction created by people to organize non-human agents. Neutral genders and the absence of gendering may also occur, but almost no research has explored this empirically in HRI or other domains in which artificial agents are studied, including HAI and CHI [35]. Alternatively, due to a discovery of male-as-default biases in HRI research [4] and the link between perceptions of agency and masculinity [22], people may perceive low anthropomorphic robots as masculine, even if not humanlike. Moreover, there may be participant-specific cultural, language, or individual cues that bias perception of gender. As such, we hypothesize that:

**H2b:** If the degree of the robot's anthropomorphism is low, people will perceive the robot to have a neutral or non-humanoid gender (e.g., gender neutral, robotic gender, no gender).

Another anthropomorphic characteristic that can intersect with gender is age. A classic study, for instance, showed that one infant given different gender labels was gendered by different adults in stereotyped ways [36]. Notably, the infant's actual gender was treated neutrally, e.g., the infant was considered genderless unless and until ascribed gender markers, such as a name. Indeed, infants and young children may colloquially be considered, or taken for granted, as "gender neutral" or "genderless." Findings for robots are few and mixed and may have been affected by methodological limitations related to the response options provided to participants. For instance, Sandygulova and O'Hare [32] provided only two options in line with the gender binary when asking children about how they would ascribe the gender and age of a robot's voice. With gender-expansive options, the results may change. As such, we hypothesize that:

**H3:** If the robot is perceived as young (baby, child), it will also be perceived as genderless.

## 3 Methods

We conducted an online within-subjects categorization study, where all participants categorized all robots, which were presented in a random order. This protocol was registered on OSF[1] before data analysis on Dec. 7th, 2021.

### 3.1 Participants

Fifty-five people located in the US were recruited through Mechanical Turk. Only those with Masters status and a HIT approval rate of over 95% were recruited. For an additional level of quality assurance, we used the procedure by Nicoletti[2], whereby we generated a set of random codes that respondents inputted on Mechanical Turk after completing the study on SurveyMonkey. Most people were aged 35 to 44 (n=21), with the rest aged 25 to 34 (n=11), 45 to 54 (n=11), 55 to 64 (n=7) and 65+ (n=5). In terms of gender, 34 identified as men/masculine and 21 identified as women/feminine; no one withheld their gender or identified as another gender, e.g., transgender, non-binary, or another gender.

### 3.2 Procedure

We used SurveyMonkey to facilitate the study. Participants were provided with a link. After giving consent on the first page, they were next presented with an overview image of the four robots in the study arranged in a circle. We did this to help generate a baseline impression of the range of robots to be categorized and thus account for issues arising from relative assessments of each individual robot. Participants were then presented with the series four robots in a random order, one per page. They were asked to carefully categorize each robot. Demographics were captured on the last page.

### 3.3 Materials

Four US-made robots were sourced from the IEEE Robots[3] database. US-made robots were chosen to match the US participant pool and thus avoid "made-in" or other effects of the robots' designs [38]. As per Spatola et al. [38], we used the Duffy taxonomy of anthropomorphism [11] to select robots representing four discrete levels of anthropomorphism: industrial, mechanical humanoid, iconic humanoid, and humanoid; figure available on OSF[4]. Like Spatola et al. [38], we cropped the images to a profile view for consistency. The robots were Cody[5], CHARLI[6], Bandit[7], and Diego-san[8].

---

[1] https://osf.io/2cuqv

[2] http://nicholasnicoletti.com/survey-monkey-and-mechanical-turk-the-verification-code

[3] https://robots.ieee.org/robots

[4] https://osf.io/duxy4

## 4 Measures

### 4.1 Perceived Gender

A custom scale based on Seaborn and Frank [34] was used to capture participants' perceptions of robot sex/gender. A 5-point gender binary scale, with feminine and masculine poles plus a neutral centre, was combined with two nominal options: "robotic-specific gender" (however perceived by each respondent) and "it does not have any sex/gender."

### 4.2 Perceived Age

A custom 5-point ordinal scale with baby (~2), child (3-12), teenager (13-19), adult (20-64), and older adult (65-) options was provided, plus an "ageless" option.

### 4.3 Perceived Anthropomorphism

The 5-point semantic differential anthropomorphism subscale from the Godspeed instrument was used [2]. The Godspeed is a well-validated scale that has been translated into many languages and is often used in HRI research.

### 4.4 Demographics

Participant demographics, reported above, were collected at the end of the survey. Age response options were provided in 5-year increments starting from 18 until 75, for which the option was "75 or older." Sex/gender [17] response options were woman/feminine, neutral/non-binary/gender fluid, man/masculine, and prefer not to say/another gender.

### 4.5 Data Analysis

Perceived gender of the robot was separated by ordinal binary continuum and nominal categories. Descriptive statistics were generated as appropriate for each type of data, i.e., averages for the continuums and counts for the nominal data. The items in the perceived anthropomorphism (Godspeed) scale were averaged together to create one score. To answer the hypotheses, we used inferential statistics, mostly relying on nonparametrics given the largely nominal data.

## 5 Results

We present the results by hypothesis below.

---

[5] https://robots.ieee.org/robots/cody

[6] https://robots.ieee.org/robots/charli

[7] https://robots.ieee.org/robots/bandit

[8] https://robots.ieee.org/robots/diegosan

## 5.1 Gendering Robots in Diverse Ways (H1)

Each robot was categorized using the range of gender-expansive options made available to participants; see Table 1. No one robot was ascribed only one gender option. We can thus accept the hypothesis that people tend to gender robots in diverse ways regardless of the robot's anthropomorphism.

**Table 1. Gender-expansive categorization of the robots**

| Robot | Anthro. | Gender Binary | | | | | Robotic Gender | No Gender |
|---|---|---|---|---|---|---|---|---|
| | | Mean | St. Dev. | Median | IQR | Count | Count | Count |
| Cody | Ind. | 1.9 | .9 | 2 | 2 | 19 | 31 | 5 |
| CHARLI | Mech. Hu. | 2.1 | .9 | 2 | 2 | 32 | 16 | 7 |
| Bandit | Ico. Hu. | 3.1 | 1.1 | 3 | 3 | 42 | 7 | 6 |
| Diego-san | Hu. | 1.3 | .6 | 1 | 3 | 52 | 1 | 2 |

## 5.2 Anthropomorphism in Perceptions of Robot Gender (H2a and H2b)

The Godspeed anthropomorphism subscale indicated a trend in perceived anthropomorphism: for Cody, a mean of 1.2 (SD=0.3, MD=1, IQR=1.3), for CHARLI, a mean of 1.6 (SD=1.6, MD=1.3, IQR=3.8), for Bandit, a mean of 2.2 (SD=0.8, MD=2, IQR=2.4), and for Diego-san, a mean of 3.5 (SD=1, MD=3.5, IQR=2.3). A significant correlation was found between these scores and our Duffy anthropomorphism categorizations, confirming our assignments, $r_s(53) = .714$, $p = < .001$.

A significant difference was also found in gender counts across the Duffy anthropomorphism levels, $X^2(6, 55) = 56.41$, $p < .001$. This indicates that gender perceptions differed by degree of anthropomorphism. The counts, for both binary and robotic genders, show an apparent trend towards an inverse relationship between degree of anthropomorphism and gender. Specifically, less anthropomorphic robots tended to be perceived as having a robotic gender, and vice versa for more anthropomorphic robots. Less anthropomorphic robots were also perceived as being more masculine. For instance, while the industrial robot, Cody, was generally ascribed a robot gender, it was also placed in the gender binary at the masculine end. However, the reverse was not true, pointing to a masculine bias. In particular, the humanoid robot, Diego-san, tended to be categorized within the gender binary at the masculine end, even though it is childlike with no gender markers. "Diego" is masculine, but this name was not revealed to participants and is not common knowledge. Still, the iconic robot (Bandit) was generally assigned a gender, one within the binary and close to the neutral centre (M=3.1, SD=1.1). This suggests that it may generally be perceived as a gender neutral or gender ambiguous robot.

Overall, we can accept the hypothesis that people tended to perceive robots with high anthropomorphism as gendered. However, we can only partially accept the hypothesis that people tended to perceive robots with low anthropomorphism as having a robotic or neutral gender. Indeed, participant ascriptions of gender for all robots indicate that these hypotheses are limited. As our analyses show, there is a complicated relationship between anthropomorphism and the gender binary. The results for the industrial robot—the least anthropomorphic—indicate diverse binary gendering when it takes place. The results for the humanoid robot will be further teased out in the next section.

## 5.3 Age in Gender and Anthropomorphism (H3)

Ascriptions of age tended to follow the robot's anthropomorphism level, with age categories assigned to more anthropomorphic robots and the "ageless" category assigned to less anthropomorphic robots; see Table 2. A significant difference was found in age category counts across the Duffy anthropomorphism levels, $X^2(6, 55) = 54.35$, $p < .001$. As with gender, the count distributions suggest an inverse relationship between perceived age and anthropomorphism.

**Table 2. Age categorization of the robots**

| Robot | Anthro. | Gender | Age | | | | | Ageless |
|---|---|---|---|---|---|---|---|---|
| | | | Mean | St. Dev. | Median | IQR | Count | Count |
| Cody | Ind. | Robot-Masculine | 4.1 | .4 | 4 | 1 | 19 | 36 |
| CHARLI | Mech. Hu. | Robot-Masculine | 3.9 | .3 | 4 | 0 | 32 | 23 |
| Bandit | Ico. Hu. | Neutral-Ambiguous | 3.5 | .8 | 4 | 2 | 42 | 13 |
| Diego-san | Hu. | Masculine | 2.1 | .4 | 2 | 2 | 54 | 1 |

Most robots (3 out of 4 or 75%) were assigned an "adult" age range (score of 3 or 4) and one was assigned a "child" age (Diego-san). For this hypothesis, we focus on the childlike robot, Diego-san, which was also categorized by ourselves and participants as humanoid, or highly anthropomorphic. As noted above, Diego-san tended to be perceived by participants as masculine. Only one participant perceived Diego-san as genderless. Based on these results, we must reject the hypotheses that robots perceived as childlike are also perceived as genderless.

After these unexpected results, we conducted an exploratory analysis of the robots based on ascribed age and gender. A significant relationship was found between gender and age across the robots, $\Phi = .60$, $p < .001$. The descriptives provide insight about this relationship. Cody, the industrial robot, was generally ascribed as "ageless" and otherwise categorized as adult in age. It also received the largest portion of "robot gender" ascriptions, while also tending towards masculine when gendered in the binary. The next robot on the anthropomorphism scale, CHARLI, had similar results: robot-masculine, adult, and otherwise roughly equal parts aged and ageless. Bandit was also generally perceived as adult in age, but neutral or ambiguous in gender. Overall, these results indicate that high anthropomorphism is associated with perceptions of age (as opposed to agelessness), as well as polar gender binary ascriptions, specifically the masculine pole. In contrast, low anthropomorphism seems to be associated with agelessness and adult agedness, as well as a mix of robot, masculine, and neutral or ambiguous gender attributions.

## 6 Discussion

A gender-expansive approach to gendering robots appears to be not only feasible but illuminating. Participants made use of the range of gender response options made available to them—notably, not just those within the gender binary. As expected, anthropomorphism influenced gendering, but in slightly unexpected ways. Notably, corollaries between binary gender options and robot gender were found. The more mechanical robots were ascribed Robot-Masculine genders, while the iconic robot was deemed Robot-Neutral and the humanoid robot was deemed Masculine (and not generally ascribed a robot gender). Gendering was not as predictable or consistent, confirming recent calls for the use of gender diverse response options and alternative genders, especially robotic or mechanical ones [34,35,37], as a critical component of research on robot gender/ing. Consensus among participants about each robot's gender was also not achieved, as the standard deviations and interquartile ranges indicate. In short, providing a set of gender-expansive response options seems to influence participant categorization of robot gender, at least with this selection of robots. Future work will need to expand the range of robots categorized to be sure.

Findings indicate that our gender-expansive framing may need further expansion. We provided a gender binary scale with a "neutral" (so-labelled) centre, as well as robotic gender and genderless options, in line with Seaborn and Frank [34]. However, the results for the iconic humanoid robot Bandit point to a limitation in this framework. Bandit's scores averaged to the middle of the gender binary or inclined to a robotic gender ascription. We interpret this to mean that Bandit was *gender ambiguous*. The response options we provided were insufficient for drawing conclusions on how exactly it was gendered. It may be gender neutral (truly genderless or without gender), *and/or* gender blended (a mix of masculine and feminine, and potentially other gender orientations), *and/or* gender ambiguous (where people could not decide on how to frame its gender) [35,41]. Moreover, it is unclear why this robot, Bandit, was an exception. Future work will need to explore the reasons directly, such as through qualitative reports from participants explaining their reasons or process of gendering. Determining whether the robot is perceived as gender neutral, blended, and/or ambiguous may also be revealed through studies placing it within different gendered and non-gendered contexts of use or interactive activities. These findings also point to potentially meaningful distinctions between *neutrality*, *ambiguity,* and *fluidity* that future theoretical work can consider more deeply and empirical work with larger samples can pinpoint. This also highlights the limitations of relying on a frame comprised of a spectrum between two poles. Categorizing complex phenomena is always tricky, as Alvin Stirling showed with respect to complexity more generally [40]. As he argues, we should avoid oversimplifying and "keep it complex" by embracing the risk, ignorance, uncertainty, and ambiguity that characterizes knowledge production. We simply remain accountable for and moveable on our decisions.

Exploring gender neutrality may be crucial in light of another major finding: the male bias effect appears to have borne out, despite the gender-expansive frame provided to participants. Yet, as Seaborn and Frank [34] argue, gender is not static. Indeed, robots and other artificial agents provide an opportunity for creative explorations of gender/ing, including shifting notions of gender for people. Understanding when and how gendering occurs in the first step. Limited and biased ascriptions of gender and any resulting effects, especially stereotyped responses, may be malleable with the right research tools, gender-conscious design, and sufficient time. Alternatively, Martin and Slepian [22] suggest that we can "de-humanize gender." Rather than "de-gendering" humans, and by extension de-gendering robots, the focus is on eliminating or complicating the idea that gender is only for humans. We can, for instance, first empirically demonstrate the absurdity of gender when applied to non-humans, and then spark questioning of these absurdities in relation to gendering people. This idea is premised in a wealth of research linking certain anthropomorphic traits, especially the Big Two of agency and communality [15,22], with the gender binary. But this is a "chicken or the egg" situation: the direction of influence is unknown. Moreover, this is the domain of social construction, which conscious and concerted effort may be able to shape. The question is then philosophical: how do we want to gender (or not)?

Age also appeared to intersect with gender, especially for the humanoid robot, Diego-san, which was designed to be and generally ascribed in this study as childlike. Unexpectedly, in consideration of human models of infants presumed to be genderless or at least malleable in their gender presentation [36], Diego-san was almost universally categorized within the binary, as masculine. The reason why needs to be explored. As discussed, the male bias effect is one possibility. Another is that there may have been unintentional gender cues in Diego-san's image. For instance, the green-and-yellow checkered shirt or metal plate on the robot's head may have been interpreted as masculine in the absence of other gender cues, as per the male bias effect. Future research with other childlike robots, including non-humanoid childlike robots, may be able to tease out how much is the male bias effect, unintentional gender cues, or other latent causes.

## 6.1 Limitations

As a pilot study, our sample size was relatively small; it will be increased in the full study to achieve greater statistical power. We were not able to provide an input field to allow participants to provide alternative gender options. This was due to a "total" question" limitation in the survey platform, on top of this research being folded into a larger project. The full study and other future work can explore open-ended gender categorization. We also evaluated only four robots, but this will be expanded in the full study. Also, given the potential of cultural differences in gender/ing, future work will need to include participants outside of the US and robots made from other countries. The full study will also involve more rigorous comparative inferential statistics, e.g., repeated-measures ANOVAs. Finally, as with other categorization research, this methodology assesses perceptions through a passive activity, which may be affected by later interaction. Future work will need to explore the effects, if any, of first impressions while also including pre-tests of gendering.

## 7 Conclusion

When it comes to robots, gender is not a given, and neither is the gender binary. As our findings show, participants ascribed a range of gender options to the robots under study. While a bias for the gender binary and in particular masculine ascriptions of robots was discovered, other options, including robotic genders no gender ascriptions, were found. Moreover, the degree of anthropomorphism and perceived age of the robot appears to intersect. Considering these findings, we argue for the use of gender-expansive response options when categorizing perceptions of robot gender. We strongly encourage those relying on perceptions of robot gender as a locus of study to conduct pilot tests and/or manipulation checks to discover how participants gender and whether it matches expectations. Future work can explore the extent to which these findings apply to other robots. The results for age and anthropomorphism suggest a path for explorations of gender neutrality, as well. Let us not take gender/ing for granted any longer.

## ACKNOWLEDGMENTS

This research was financially supported by the university department of the first and last author.